\documentstyle[aps,prl,multicol]{revtex}
\begin{document}
\def\be{\begin{equation}}
\def\ee{\end{equation}}
\def\bc{\begin{center}}
\def\ec{\end{center}}
\def\bea{\begin{eqnarray}}
\def\eea{\end{eqnarray}}

\title{The approach to scaling in phase-ordering kinetics}
\author{C. Castellano$^{1}$ and M. Zannetti$^{2}$}
\address{$1$ Dipartimento di Scienze Fisiche, Universit\`{a}
di Napoli, Mostra d'Oltremare, Padiglione 19, 80125 Napoli, Italy}
\address {$2$ Istituto Nazionale di Fisica della Materia, Unit\`{a}
di Salerno and Dipartimento di Fisica, Universit\`{a} di Salerno,
84081 Baronissi (Salerno), Italy}
\maketitle
\begin{abstract}
The influence of the initial fluctuations on 
the onset of scaling in the quench to zero temperature
of a two dimensional system with conserved order parameter, 
is analyzed in detail with and without
topological defects. 
We find that the initial fluctuations greatly
affect the way scaling is reached, while the number
of components of the order parameter does not play a
significant role. Under strong initial fluctuations
the preasymptotic linear behavior is replaced by the mean field behavior
of the large-$N$ model, producing 
the multiscaling to standard-scaling crossover also in 
the physically interesting cases of systems with a small number 
of components. 
\end{abstract}
PACS:  64.60.Cn, 05.70.Fh, 64.60.My, 64.75.+g

\begin{multicols}{2}
Much of the interest in phase ordering-kinetics is devoted to
scaling behavior of the structure factor~\cite{Bray94}.
This is a late stage phenomenon which takes place after the order
parameter has reached local equilibrium and the typical domain
size $L(t)$ dominates all other lengths. 
The structure factor obeys the scaling form
\be
C({\vec k},t) = L^d(t) F(x)
\label{SS}
\ee
where $x = kL(t)$, $L(t) \sim t^{1 \over z}$ and,
when the order parameter is conserved,
$z=3$ or $z=4$ in the scalar or vectorial case, respectively.
All what happens before, in the time span between the instant of
the quench and the beginning of the late stage, is generically referred
to as early stage.
Usually the early stage theory is identified with the Cahn-Hilliard
linear theory~\cite{Cahn59} and early stage 
studies are mainly focused on how and 
when this theory breaks down~\cite{Binder84,Elder88,Laradji90,Corberi95}.
From investigations along this
line emerges a picture of the early stage as a time regime where 
quite complex and interesting processes take place. In particular, it
appears that the breakdown of the linear theory starts first at
short distances and propagates with time towards larger length 
scales~\cite{Gross94}.

Recently~\cite{Castellano96}, the early evolution 
of the phase-ordering process with
a conserved order parameter has been investigated in a different
context with the aim of investigating the multiscaling to
standard-scaling crossover. In the exact solution of the large-$N$
model~\cite{Coniglio89} the late stage form (\ref{SS}) is replaced by the more
general multiscaling behavior
\be
C({\vec k},t) = \left[k^{2-d}_m(t) L^2(t)\right]^{\varphi(x)} F(x)
\label{MS}
\ee
where $k_m(t)$ is the peak wave vector, $x = k/k_m$, 
$\varphi(x)= \psi(x)$ with $\psi(x) \equiv 1-(x^{2}-1)^{2}$ 
and $k_{m}L(t) \sim (\ln t)^{1/4}$.
It was shown by Bray and Humayun~\cite{Bray92} (BH) 
in the framework of the $1/N$
expansion that (\ref{MS}) holds as a late stage property only when $N$, 
the number of the order parameter components, is strictly infinite
while the standard-scaling form (\ref{SS}) is eventually restored
for any finite value of $N$. The full
time evolution of the structure factor for $N$ large but finite is then
expected to display the crossover from multiscaling 
to standard-scaling, with a crossover time which grows
with $N$ and also with the size $\Delta$ of the fluctuations in the
initial state. Indeed, the existence of the crossover together with
this dependence of the crossover time on $N$ and $\Delta$ has been
found in the numerical solution~\cite{Castellano96} of the BH model. 

This result suggests to investigate the early
behavior of systems with a small number of components and large
initial fluctuations, for which no early stage theory is available.
In this Letter we will focus on how the size of the fluctuations in
the initial condition affects the approach to scaling.
The influence of the initial fluctuations on the structure of the
early stage was previously pointed out by Elder, Rogers and Desai
~\cite{Elder88}.
Here we are able to make a detailed analysis of this effect
and to complete the picture of what happens before scaling is
fully developed.

We consider the zero temperature quenches of a two-dimensional
system described by the Cahn-Hilliard equation
\be
{\partial {\vec \phi} \over \partial t} =
- \nabla^2 [\nabla^2 {\vec \phi} - V'(\phi)]
\label{CH}
\ee
with
$V({\vec \phi}) = {r \over 2} {\vec \phi}^{2} + 
{g \over 4} ({\vec \phi}^2)^{2}$
and $r <0$, $g >0$ in the two cases of $N=1$ (singular topological 
defects) and $N=4$ (absence of topological defects).
In the initial state the system is uncorrelated
\be
<\!\!{\vec \phi}({\vec x}) \cdot {\vec \phi}({\vec x}')\!\!>
= \Delta \delta({\vec x} - {\vec x}') 
\ee
and $\Delta$ measures the strength of the fluctuations.
We will consider the two cases
$\Delta \ll {\vec \phi}_{eq}^2$ (small $\Delta$)
and $\Delta \simeq {\vec \phi}_{eq}^2$ (large $\Delta$), where 
$\left|{\vec \phi}_{eq}\right| = \sqrt{-r \over g}$ is the
equilibrium value of the order parameter.
We shall find that when $\Delta$ is large the early 
stage of the phase-ordering process is 
accurately described by the large-$N$ model 
(mean field theory), both for $N=1$ and for $N=4$. 
Then, the multiscaling to standard-scaling
crossover is expected to be a generic feature of the early stage, when the order
parameter is conserved and fluctuations in the initial state
are large. Furthermore, we shall see that the structure of the
crossover follows a pattern similar to what one has in the
transition from linear to nonlinear regime when the initial
fluctuations are small~\cite{Gross94}. Namely, it turns out that also the
mean field theory breakdown takes place through the propagation
of the failure from the short to the large length scales.

The tool which allows to monitor all possible
behaviors is the measurement of the spectrum of exponents $\varphi(x)$
in (\ref{MS}) over a sequence of time intervals~\cite{Castellano96}. 
The form (\ref{MS}) contains (\ref{SS}) as a particular case
with $\varphi(x) \equiv 1$. Considering
$C({\vec k},t)$ in the time interval ($t_0$,$\tau$) beginning at $t_0$
and of duration $\tau$, after taking the logarithm of Eq.(\ref{MS}) in two
dimensions we have
\be
\ln C({\vec k},t) = \varphi({\vec x}) \ln L^{2}(t) + \ln F(x)
\label{logMS}
\ee
which yields $\varphi(x)$ as the slope of 
$\ln C({\vec k},t)$ vs $\ln L^{2}(t)$,
over the interval ($t_0, \tau$), at fixed $x$.
In principle, $\varphi(x)$ may depend on $t_0$.
If this happens scaling does not hold. Instead, if $\varphi(x)$
does not depend on the time interval there is scaling and more
specifically there is multiscaling if $\varphi(x)$ depends on $x$, while
scaling is standard if $\varphi(x)$ is independent of $x$.

The time evolution of the structure factor has been obtained numerically
by simulating the equation of motion with a computationally
efficient cell dynamical system~\cite{Oono87}.
The details of the computation are the same of Ref.~\cite{Mondello93}.
We used values of the parameters such that that ${\vec \phi}^2_{eq} = 0.9558$
and the initial amplitudes of the
fluctuations were $\Delta = 2.083$ (large $\Delta$) or
$\Delta = 8.3 \times 10^{-8}$ (small $\Delta$).
The results were averaged over 30 realizations of the initial conditions
on a $256 \times 256$ lattice for $N=4$ (small and large $\Delta$) and
$N=1$ (small $\Delta$). For $N=1$ and large $\Delta$ we used a
$512 \times 512$ lattice and averaged over 90 realizations (up to $t= 5432$)
and over 50 realizations (up to $t= 54177$).

Let us first go over the case of small $\Delta$.
In the behavior of $C({\vec k},t)$
for $N=1$ (Fig.~\ref{1}) one can recognize two sharply separated time 
regimes, one with $k_m$ constant and the other with $k_m$
decreasing with time (see also Fig.~\ref{5}). 
The latter corresponds to the late stage
which displays coarsening and 
a well formed $k^{-3}$ Porod tail for $k>k_{m}$,
indicating the existence of sharp interfaces. 
For $k < k_{m}$  
the structure factor displays the $k^{4}$ power law
~\cite{Yeung88}. Interesting non trivial behavior can be
observed also in the early stage  where $k_{m}$ is
constant. Here two subregimes can be identified.
At the very beginning, as we shall see more precisely below, the
evolution is linear except for the high wave vector region where 
an incipient Porod tail appears. Later on, but with $k_{m}$ still time
independent, the evolution is nonlinear for all $k$'s due to the 
appearence of the power law behavior also for $k<k_{m}$. 
The evolution takes place exactly in the same way for $N=4$,
except that the Porod tail is replaced by exponential decay,
due to the absence of topological
defects~\cite{Castellano96,Rojas96}.

Next, in order to probe the scaling behavior, we have measured
$\varphi(x)$ in a sequence of seven time intervals (Fig.~\ref{2}).
The first four curves display a strong dependence of $\varphi(x)$
both on $x$ and on the time interval, while the remaining three,
for $t \geq 500$, are essentially flat and time independent. Accordingly,
the latter ones belong to the late stage with standard-scaling already
established, while the set of the first four clearly illustrates
the absence of scaling in the early stage. Furthermore, a
closer analysis of the $x$ dependence of these curves provides
evidence, also in the case with conserved order parameter, for
the propagation of the breaking of linear evolution from
high towards low wave vectors, as previously observed
in the nonconserved case~\cite{Gross94}. In fact, these curves display
a minimum which moves towards smaller values of $x$ as time 
proceeds and the behavior of $\varphi(x)$ to the left of the minimum
can be accounted for on the basis of the linearized theory, but not
on the right.
When $\Delta$ is small
at the very beginning the non linear term can be
neglected in Eq.(\ref{CH}) and the analytical form of the structure 
factor can be written as
\be
C({\vec k},t) = \Delta \exp\left[2 k_m^4 t \psi(x)\right]
\ee
with $k_m^2 = - r/2$ and $x = k/k_m$.
Then, taking the logarithm
\be
\ln C({\vec k},t) = \left[{2 k_m^4 t \over \ln L^{2}(t)} \psi(x)\right]
\ln L^{2}(t) + \ln \Delta
\ee
and comparing with Eq.(\ref{logMS}) we can identify
$\varphi(x) = {2 k_m^4 t \over \ln L^{2}(t)} \psi(x)$. Indeed, the behavior of
$\varphi(x)$ to the left of the minimum for curves up to
$t \sim 200$ is well fitted by $\psi(x)$  with a growing time dependent
prefactor. The behavior of $\varphi(x)$ for $N=4$ is identical
except for a slight change in the vertical scale.

The picture changes completely when $\Delta$ is large. The time evolution
of $C({\vec k},t)$ is displayed in Fig.~\ref{3} and the main qualitative
differences with Fig.~\ref{1} are that 
there is no regime with constant $k_{m}$ (see also Fig.~\ref{5})
and that it takes much longer to build up the $k^4$ tail for
$k<k_m$. 
But the real insight into what is going on is obtained from
the behavior of $\varphi(x)$.
In Fig.~\ref{4} and \ref{4bis} we have plotted the time 
evolution of $\varphi(x)$ 
over six different time intervals. Again $\varphi(x)$ displays a
minimum which moves towards lower values of $x$ as time goes on.
However, the behavior is much different from what we had with
small $\Delta$. Considering the set of the first four
curves for $N=1$ (Fig.~\ref{4}), 
while on the right of the minimum there is no scaling as before, 
now on the left there is multiscaling with $\varphi(x)$
time independent and well described by $\psi(x)$
(notice that these curves are magnified by factors of
30 with respect to those in 
Fig.~\ref{2}).
For larger times (curves from $t \sim 1700$ onward) the
opposite occurs, namely on the right of the minimum $\varphi(x)$
tends to lose the time dependence as well as the $x$ dependence,
while on the left there appears a time dependence. This
overall behavior is quite informative. The set of the first four curves
shows that in the early stage mean field theory provides an
adequate treatment of the nonlinearity all the way down to a
certain length scale (corresponding to the minimum in $\varphi(x)$),
which grows with time. The description of what happens below this cutoff
is beyond the reach of mean field theory. Later on ($t \geq 1700$)
mean field theory breaks down over all length scales, since
the appearance of a flat 
behavior together with the disappearance of the time dependence
for $x \geq 1$ signals the onset of standard-scaling on the shorter distances,
while the time dependence  for $x \leq 1$  shows that 
the crossover towards
standard-scaling is under way over larger distances. 
The $N=4$ (Fig.~\ref{4bis}) system behaves essentially in the
same way, except that the breakdown of multiscaling for
$x \leq 1$ does not show up in the time of the simulation,
indicating that the crossover time is larger for $N=4$ than for
$N=1$. It is worth stressing that from the above analysis
follows that the time regime preceding the fully developed
standard-scaling under strong initial fluctuations can be
made to last much longer than the corresponding early stage
when the initial fluctuations are small.

Additional insight into the difference between small $\Delta$ and
large $\Delta$ can be obtained from the behavior of the peak wave vector
(Fig.~\ref{5}). 
The curves for small $\Delta$ show clearly the existence of the early stage
with $k_{m}^{-1}$ constant followed by a sharp transition to power
law growth in the scaling regime. Instead, for
large $\Delta$ the plateau is replaced by a sharp drop of $k_{m}^{-1}$,
soon followed by increasing behavior. 
This initial decreasing behavior of $k_{m}^{-1}$ 
and the subsequent change to positive slope are consistent with
the solution of the large-$N$
model~\cite{Coniglio89}. Furthermore, the asymptotic slope is reached very slowly.
We have analyzed the growth of $k_m^{-1}(t)$
with the power law $t^{1/z}$ and the results for three time intervals
are displayed in Table 1.
The interesting feature here is that the asymptotic values $1/3$ for $N=1$ and 
$1/4$ for $N=4$ are approached from above for small $\Delta$ and from below
for large $\Delta$. Thus, in particular, for $N=1$ and $\Delta$ large
the effective growth exponent before reaching values close to
$1/3$ goes through a time region where it is closer to $1/4$.
This is consistent with the mean field description
of the early stage when $\Delta$ is large and shows that the behavior of the 
effective exponent is quite sensitive to the initial condition
for a long time.
 
In conclusion, we have analyzed the approach to scaling in the
quench to zero temperature of a two dimensional system, which
may ($N=1$) or may not ($N=4$) form stable topological defects,
under the effect of weak and strong initial fluctuations.
We have found that, for physically interesting systems,
the value of $N$ plays no significant role,
while the strength of the initial fluctuations is quite important.
When $\Delta$ is small there is a clearly identifiable and relatively
short early stage, with no scaling, dominated by the formation of
domains on a fixed length scale.
After these have been formed, coarsening begins, the late stage is
entered and dynamics obeys standard-scaling.
When $\Delta$ is large, instead, the initial process of domain formation
is bypassed and coarsening begins right away,
but it is not accompanied by standard-scaling. Rather, before
standard-scaling sets in there is a long time transient characterized
by multiscaling which shows that the early stage evolution is
governed by mean field theory. So we are now in the position
to make an assessment of what is the proper place for the large-$N$
model in the theory of phase-ordering processes for systems
with a small number of components. It is clear, on the basis
of the BH result and of more recent work of Rojas and Bray~\cite{Rojas96},
that the large-$N$ model does not give reliable information
on the late stage behavior of these systems, neither the 
$1/N$ expansion offers a viable systematic approximation
scheme. Instead, as we have found, the large-$N$ model
qualifies for the early stage theory, whenever the initial
fluctuations are large.

We thank Yoshi Oono for providing the original CDS program from
which we have developed ours. One of us (C.C.) whishes also to
thank the National Institute of Standards and Technology for
the kind hospitality during the completion of this work.

\bc
\bf FIGURE CAPTIONS
\ec

\begin{enumerate}
\item \label{1}
Time evolution of $C(\vec{k},t)$ for small $\Delta$ and $N=1$.
\item \label{2}
Time evolution of $\varphi(x)$ for small $\Delta$ and $N=1$.
\item \label{3}
Time evolution of $C(\vec{k},t)$ for large $\Delta$ and $N=1$.
\item \label{4}
Time evolution of $\varphi(x)$ for large $\Delta$ and $N=1$.
\item \label{4bis}
Time evolution of $\varphi(x)$ for large $\Delta$ and $N=4$.
\item \label{5}
Time evolution of the peak wave vector.
\end{enumerate}

\bc
\bf TABLE
\ec
\bc
\begin{tabular}{||ll|r@{.}l@{$\pm$}r@{.}l|r@{.}l@{$\pm$}r@{.}l|r@{.}l@{$\pm$}r@{.}l||} \hline
 \multicolumn{2}{||c|}{Time intervals}
 & \multicolumn{4}{c|}{274--1086}
 & \multicolumn{4}{c|}{1367--5432}
 & \multicolumn{4}{c||}{6837--54177} \\[2mm] \hline
$N=1$\hspace{5mm} small&$\Delta$ & 0&47  &0&03  &0&36 &0&01  &0&351&0&007 \\
$N=1$\hspace{5mm} large&$\Delta$ & 0&223 &0&008 &0&275&0&003 &0&322&0&004 \\
$N=4$\hspace{5mm} small&$\Delta$ & 0&42  &0&03  &0&295&0&005 &0&266&0&005 \\
$N=4$\hspace{5mm} large&$\Delta$ & 0&202 &0&008 &0&220&0&07  &0&235&0&002 \\ \hline
\end{tabular}
\ec

\noindent
Table 1. Values of the exponent $1/z$ computed by linear regression on
the data displayed in Fig.~\ref{5}.

\end{multicols}
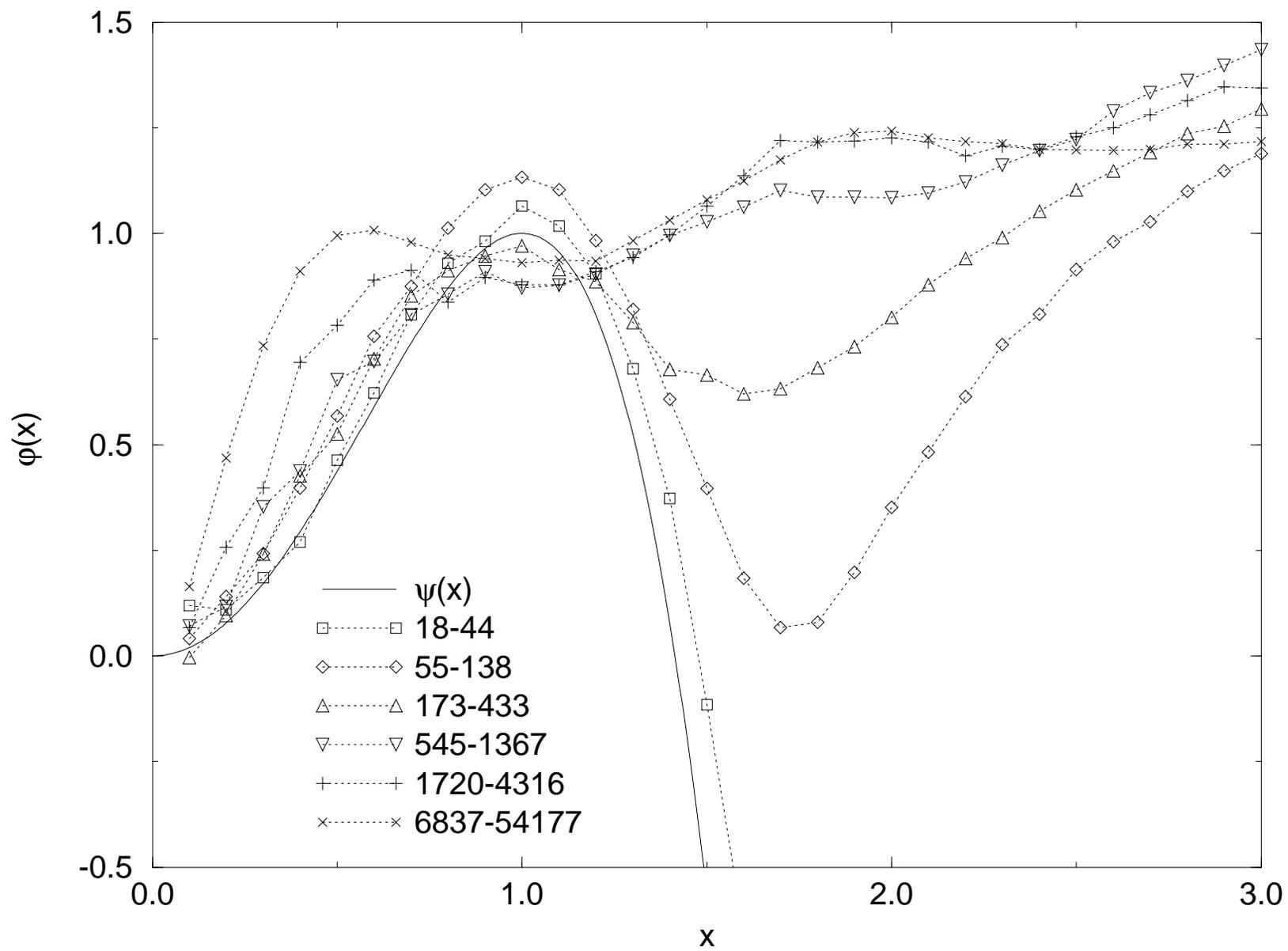
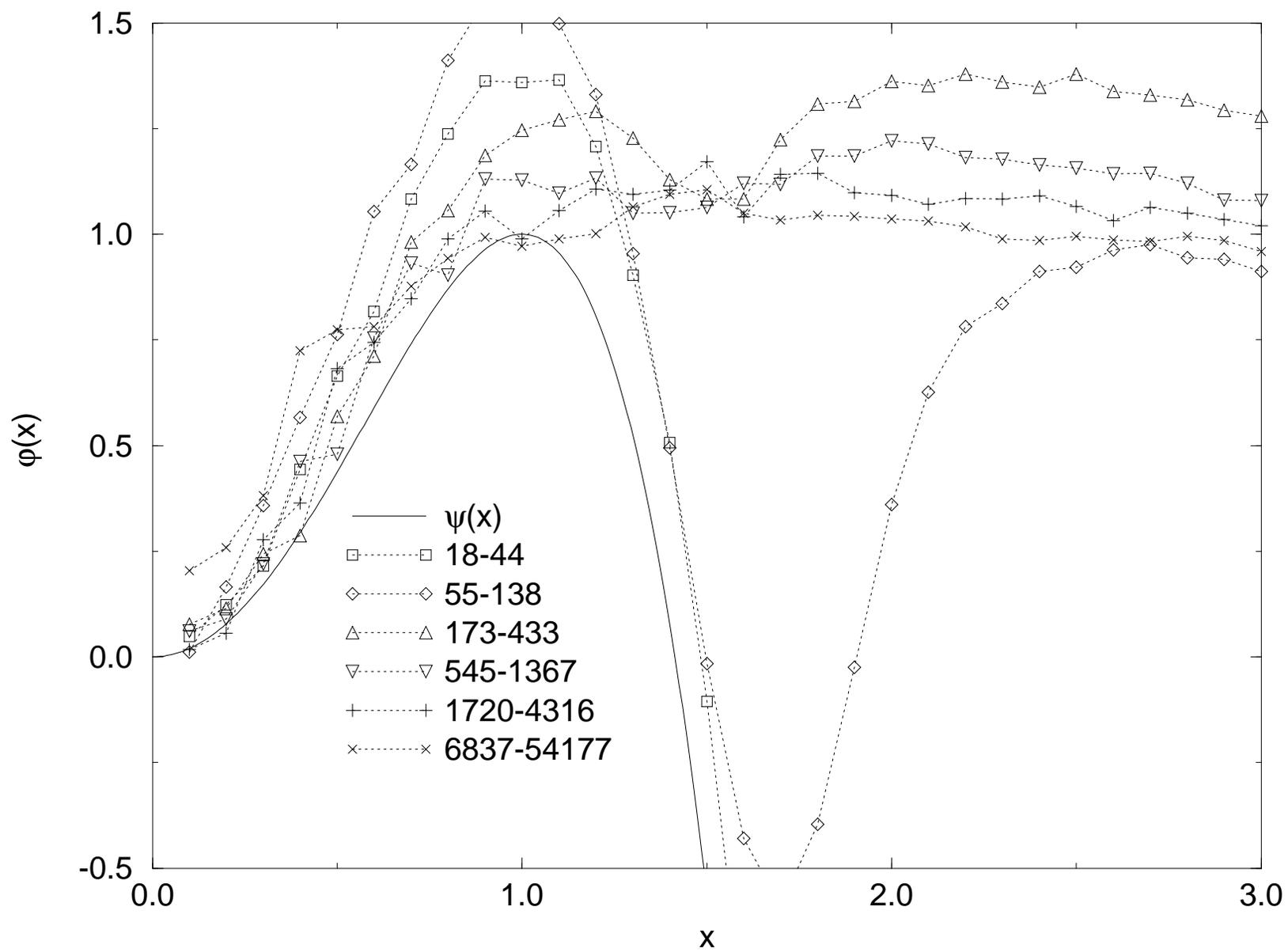
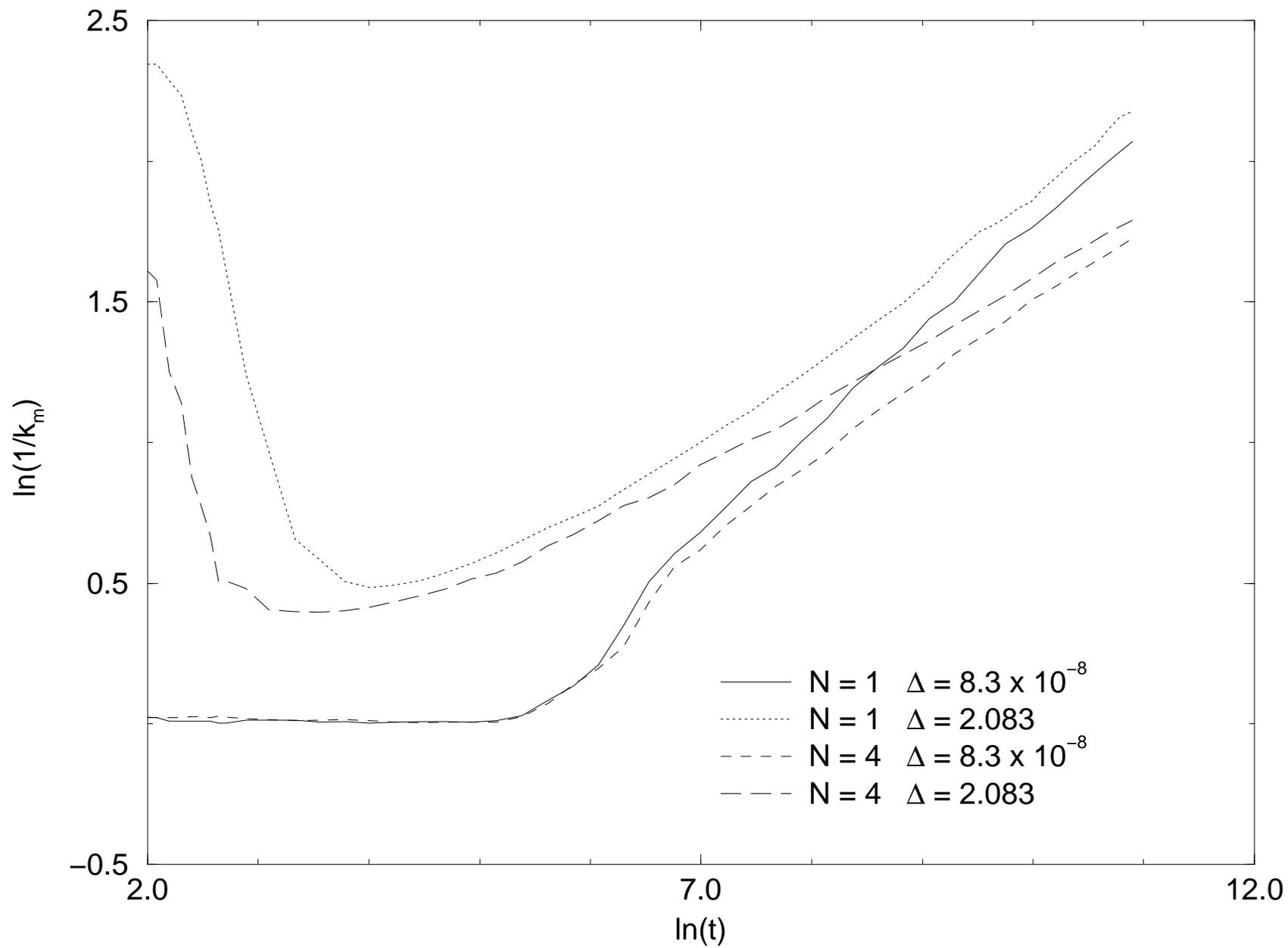
\end{document}